\documentclass{article}

\usepackage{arxiv}

\usepackage[utf8]{inputenc}
\usepackage[T1]{fontenc}
\usepackage[numbers,sort&compress]{natbib}
\usepackage{amsmath,amssymb,amsfonts}
\usepackage{graphicx}
\usepackage{booktabs}
\usepackage{microtype}
\usepackage{hyperref}
\usepackage{url}
\usepackage{doi}
\usepackage{here}

\title{Lifecycle-Based Design and Evaluation of Real-Time Backup Triggers for Ransomware Damage Mitigation}

\author{
  \href{https://orcid.org/0009-0001-0594-4620}{%
    \includegraphics[scale=0.06]{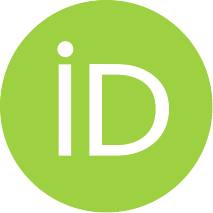}\hspace{1mm}Kosuke Higuchi} \\
  Kogakuin University\\
  1-24-2 Nishi-Shinjuku, Shinjuku-ku\\
  Tokyo, Japan\\
  \texttt{ed25001@ns.kogakuin.ac.jp}
  \And
  \href{https://orcid.org/0000-0001-5956-3455}{%
    \includegraphics[scale=0.06]{orcid.pdf}\hspace{1mm}Ryotaro Kobayashi} \\
  Kogakuin University\\
  1-24-2 Nishi-Shinjuku, Shinjuku-ku\\
  Tokyo, Japan\\
  \texttt{ryo.kobayashi@cc.kogakuin.ac.jp}
}

\renewcommand{\shorttitle}{Lifecycle-Based Real-Time Backup Triggers}

\hypersetup{
  hidelinks,
  hypertexnames=false,
  pdftitle={Lifecycle-Based Design and Evaluation of Real-Time Backup Triggers for Ransomware Damage Mitigation},
  pdfsubject={Computer Science--Cryptography and Security},
  pdfauthor={Kosuke Higuchi, Ryotaro Kobayashi},
  pdfkeywords={backup trigger timing, eBPF, ransomware mitigation, real-time backup},
}

\begin{document}
\maketitle

\begin{abstract}
Ransomware continues to encrypt files during the interval between attack onset and detection.
Real-time backups can mitigate this damage by preserving files before they are modified.
The previously proposed Real-Time Open-File Backup System (ROFBS) triggers backups primarily on file-open events.
However, the file lifecycle offers several candidate trigger points, including open, read, write, and rename operations.
Triggering backups too early may create unnecessary backup files, whereas triggering them too late may allow ransomware writes to race with backup creation and prevent the preservation of clean file contents.
Consequently, it remains unclear which trigger timing best balances recoverability and the number of backups created.
In this study, we design and evaluate real-time backup triggers for mitigating ransomware damage from a file-lifecycle perspective.
Specifically, we compare four strategies: Open-time backup, Read-time backup, Write-time backup, and Rename-time backup.
We implement these strategies in an ROFBS-style prototype on XFS and evaluate them using five ransomware samples: Conti, Sodinokibi, AvosLocker, REvil, and HelloKitty.
Our results clarify how trigger timing affects both damage mitigation and the number of backups created, providing design guidance for selecting effective triggers in real-time backup systems against ransomware.
\end{abstract}


\section{Introduction}
Crypto-ransomware encrypts files stored on compromised endpoints and servers and demands payment in exchange for the decryption key.
Such attacks can severely affect victim organizations by disrupting operations and increasing recovery costs.
Ransomware remains a major cyber threat, with incidents continuing to be reported in recent years \cite{verizonDBIR2025}.
According to CyberEdge, 72.7\% of organizations were affected by ransomware attacks in 2023, the highest percentage recorded since 2018 \cite{cyberedge}.

One factor contributing to the proliferation of ransomware attacks is the widespread adoption of the Ransomware-as-a-Service (RaaS) business model.
RaaS provides the tools and infrastructure required to conduct ransomware attacks as a service, enabling attackers without advanced technical expertise to launch attacks with relative ease \cite{raas}.
Against this background, ransomware attacks have become increasingly sophisticated and diverse.
Allianz projects that the annual cost of ransomware will reach approximately USD~265 billion by 2031 \cite{allianz}.

A file-open operation is one of the file operations commonly observed across diverse ransomware families during ransomware encryption.
Leveraging this property, we proposed the Real-Time Open-File Backup System (ROFBS) as a real-time defense against ransomware encryption attacks \cite{Higuchi2025}.
ROFBS creates a real-time backup of a target file when the file is opened.
We defined the proportion of files protected during a ransomware attack as the Backup Ratio and experimentally demonstrated that ROFBS consistently achieved high Backup Ratios.
This work is important because it demonstrated the potential to preserve files before encryption by using ransomware file operations as backup triggers.
Ransomware encryption, however, involves multiple file operations, including not only open but also read, write, and rename operations.
The order of these operations and the files on which they operate may vary with the encryption scheme employed by the ransomware.
Consequently, the operation selected as the backup trigger may affect both when backup creation begins and the proportion of files that can be protected.
Previous work evaluated only a strategy triggered by file-open operations and did not compare it with strategies triggered by read, write, or rename operations.
It therefore remains unclear which file operation is the most effective backup trigger for protecting files against ransomware.

In this study, we revisit the backup mechanism of ROFBS from the perspective of selecting the file operation used to trigger backup creation.
Specifically, we reimplement ROFBS and introduce read, write, and rename operations as backup triggers in addition to the conventional open operation.
Using multiple ransomware samples, we then compare the Encrypted Files, Backup Files, Covered Files, and Backup Ratio obtained with each trigger for the same sample.
This evaluation clarifies how the file operation selected as the backup trigger affects ransomware encryption damage and backup performance.

\section{Related Work}

This section classifies defense methods against ransomware into four categories: prevention and obfuscation, early detection and monitoring, decoy-based approaches, and recovery and storage approaches.
Representative studies in each category are reviewed below.

\subsection{Prevention- and Obfuscation-Based Approaches}

Alshaikh et al. proposed the Ransomware Prevention Toolkit (RPTK), a proactive hybrid defense framework that exploits the tendency of ransomware to exclude system files from encryption in order to keep the compromised host operational \cite{Alshaikh2026}.
In an evaluation using 50 samples selected from major crypto-ransomware families, RPTK detected and terminated all 50 samples within the scope of the evaluation, achieving a detection accuracy of 100\% according to the authors' definition.

McIntosh et al. proposed Staged Event-Driven Access Control (SEDAC), which combines program-centric and user-centric access control \cite{McIntosh2023}.
Experiments using a prototype implemented on Windows demonstrated that SEDAC could block a broader range of ransomware attack vectors than existing methods.

Lee et al. proposed a method for protecting files against existing ransomware by applying a concealment strategy that makes important files more difficult to discover when ransomware searches for files to encrypt \cite{Lee2023}.
Experiments using real ransomware samples demonstrated that the proposed method could protect important files at a low cost.

\subsection{Early Detection and Monitoring Approaches}

Zhang et al. proposed Ranker, a real-time method for early ransomware detection based on kernel-level behavior analysis \cite{Zhang2024}.
Ranker analyzes kernel-level file I/O and process behavior before and after the onset of mass encryption.
In an evaluation covering 68 ransomware families, it achieved an average F1 score of 99.43\% and a false-positive rate of 0.11\%.

Kim et al. proposed two byte-frequency-based metrics, EntropySA and DistSA, that use the characteristics of sample areas (SAs), which are specific subregions within files, to detect file encryption performed by ransomware \cite{Kim2022}.
Experiments showed that the proposed metrics achieved an F1 score of 0.994 and an average detection rate of 99.46\% for ransomware file-encryption attacks.

Lee et al. proposed Rcryptect, a system that uses FUSE to monitor data blocks in the file system and detect encryption operations in real time \cite{Lee2022}.
Experiments demonstrated that Rcryptect could detect and prevent encryption by known ransomware, identify multiple encryption operations, and limit performance overhead to approximately 13\%.

Kok et al. proposed the Pre-Encryption Detection Algorithm (PEDA), which uses machine learning to analyze patterns of Windows API calls made by suspicious programs and detect crypto-ransomware before file encryption begins \cite{Kok2019}.
In experiments comparing PEDA with Naive Bayes, Random Forest, an ensemble of these two methods, and EldeRan, PEDA achieved the lowest false-positive rate, at 1.56\%.

Berrueta et al. proposed a machine-learning-based method for detecting and blocking crypto-ransomware by extracting features related to file reads and writes from file-sharing communications between clients and file servers \cite{Berrueta2022}.
The method can also operate when the communication traffic is encrypted.
Experiments demonstrated that the proposed method detected all of the more than 70 ransomware samples evaluated, including previously unseen samples that were not used for training.

Lee et al. proposed a machine-learning-based method for detecting ransomware-infected files encrypted using format-preserving encryption (FPE), which suppresses changes in file formats and entropy to evade conventional entropy analysis \cite{Lee2025}.
Experiments across multiple datasets achieved an average precision of 94.64\% and demonstrated that many machine-learning models could effectively detect infected files encrypted using FPE.

\subsection{Decoy-Based Approaches}

Sheen et al. proposed R-Sentry, a lightweight real-time detection method that uses honeyfiles as decoys to detect crypto-ransomware at an early stage \cite{SHEEN2022}.
By analyzing file-search patterns employed by existing and future ransomware variants and determining appropriate honeyfile placement across multiple folders, R-Sentry can rapidly detect ransomware that follows the identified search patterns.

Gómez-Hernández et al. proposed R-Locker, a Windows-based tool that detects and blocks crypto-ransomware at an early stage by placing numerous decoy files in the file system and monitoring access to them \cite{Gomez2022}.
Experiments demonstrated that the Windows version of R-Locker could effectively and efficiently prevent file encryption by ransomware while minimizing its impact on normal system operation.

\subsection{Recovery- and Storage-Based Approaches}

Dafoe et al. proposed FFRecovery, a defense method that enables fine-grained, file-level data recovery after ransomware attacks \cite{Dafoe2024}.
FFRecovery reconstructs metadata through file-system forensics, extracts raw data from the Flash Translation Layer (FTL), and combines the reconstructed metadata and extracted data to recover files.
Experiments using real ransomware samples demonstrated the effectiveness of the proposed method.

We proposed the Real-Time Open-File Backup System (ROFBS), which immediately creates a backup when a file is opened to minimize the encryption damage that occurs between ransomware detection and termination \cite{Higuchi2025}.
Experiments demonstrated that ROFBS maintained a high Backup Ratio during ransomware attacks while imposing only a small impact on CPU usage, memory usage, and disk I/O during normal operation.

Kim proposed a method that detects encryption damage in real time by analyzing the fixed structures and header information of files and recovers files from temporary copies retained before I/O processing \cite{Kim2025}.
Experiments showed that Simple Format Analysis (SFA) achieved detection accuracies of 95.0\% with SFA-F and 97.9\% with SFA-F-H.
SFA-F-H, which combines fixed structures and header information, achieved the highest detection accuracy.
\section{Background}
\label{sec:background}

\subsection{Extended Berkeley Packet Filter}
\label{subsec:ebpf}

Extended Berkeley Packet Filter (eBPF) is a technology that enables user-defined programs to execute in kernel space in response to events within the Linux kernel \cite{eBPF}.
eBPF enables kernel operations to be observed without directly modifying the kernel source code or developing custom kernel modules.
Before an eBPF program is loaded into the kernel, it is validated by the verifier, which restricts operations that could compromise kernel safety, such as invalid memory accesses and non-terminating execution.
Owing to these characteristics, eBPF is widely used for tracing, performance analysis, network processing, and security monitoring.

\subsection{BPF Compiler Collection}
\label{subsec:bcc}

BPF Compiler Collection (BCC) is a toolkit that facilitates the development of tracing and monitoring programs based on eBPF \cite{bcc}.
BCC allows developers to write kernel-space eBPF programs in C and implement the user-space components that load those programs and process the collected data in Python or Lua.
kprobes and kretprobes are used to observe kernel functions.
A kprobe is triggered when a target function is invoked and provides access to information such as its arguments.
In contrast, a kretprobe is triggered when the target function returns and provides access to its return value and results.

\section{Backup Trigger Design}
\label{sec:backup_trigger_design}

We construct a real-time backup mechanism that creates backups in response to file operations.
The original ROFBS design begins backing up a target file when it detects an open operation on that file.
Ransomware encryption, however, involves multiple file operations, including open, read, write, and rename operations.
Because these operations occur at different points in the file lifecycle, the likelihood of preserving pre-encryption content may depend on which operation triggers backup creation.

Ransomware may also encrypt files using different file-operation sequences.
To illustrate how these differences affect backup trigger timing, we introduce two simplified conceptual models.
These models are not intended to reproduce the complete behavior of any particular ransomware sample.
Instead, they illustrate the relative timing of the candidate backup triggers and differences in the files on which the operations are performed.
The circles in the figures indicate the open, read, write, and rename trigger points evaluated in this study.

\begin{figure}[t]
  \centering
  \includegraphics[width=\columnwidth]{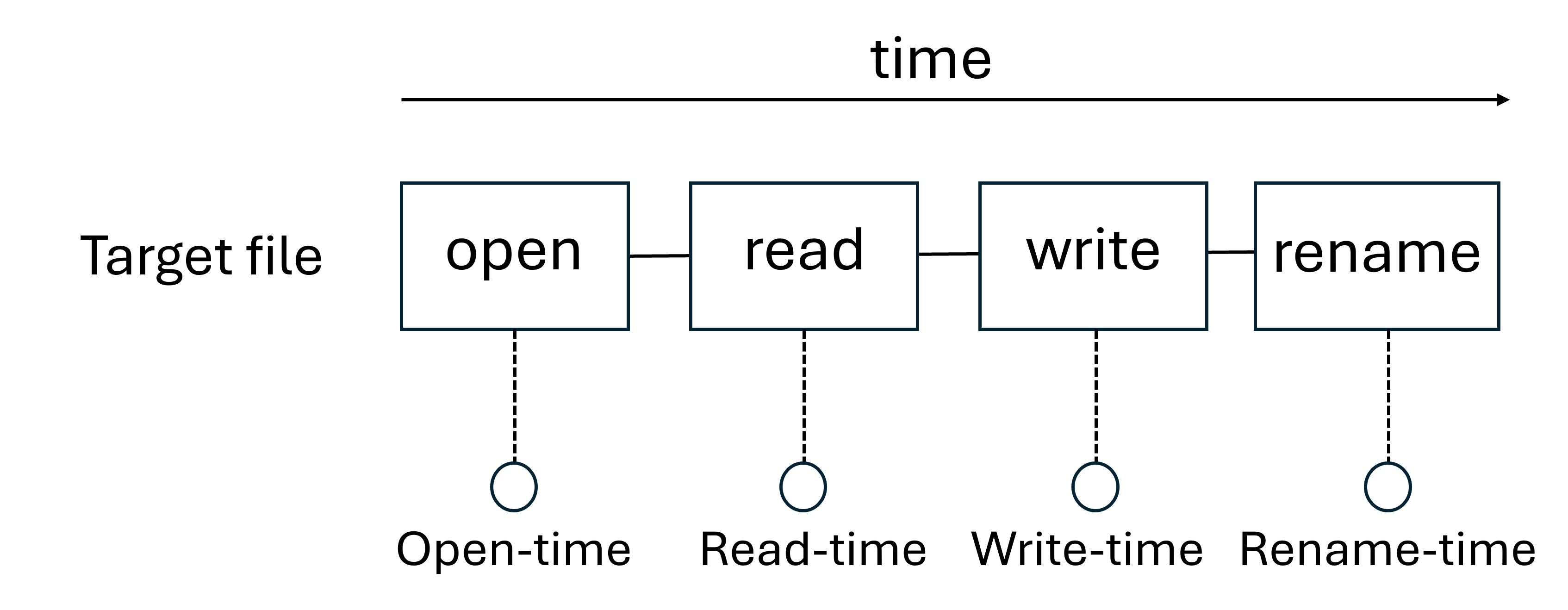}
  \caption{Conceptual model of in-place encryption and candidate backup trigger positions.}
  \label{fig:inplace_encryption}
\end{figure}

Fig.~\ref{fig:inplace_encryption} shows an in-place encryption model.
In this model, ransomware opens and reads an original file, writes encrypted content back to the same file, and may subsequently rename the file.

In the in-place encryption model, the open and read operations occur before the encrypted content is written.
Therefore, strategies triggered by open or read operations may begin backup processing earlier than a strategy triggered by a write operation.
With a write trigger, the ransomware's encryption write and the backup process may proceed concurrently.
Consequently, part or all of the original file may be modified before the backup process is completed.
If a rename operation occurs after encryption, the file content may already have been modified by the time the rename operation is detected.
A rename operation is optional in this model because not all in-place encryption procedures rename the encrypted file.

\begin{figure}[t]
  \centering
  \includegraphics[width=1\columnwidth]{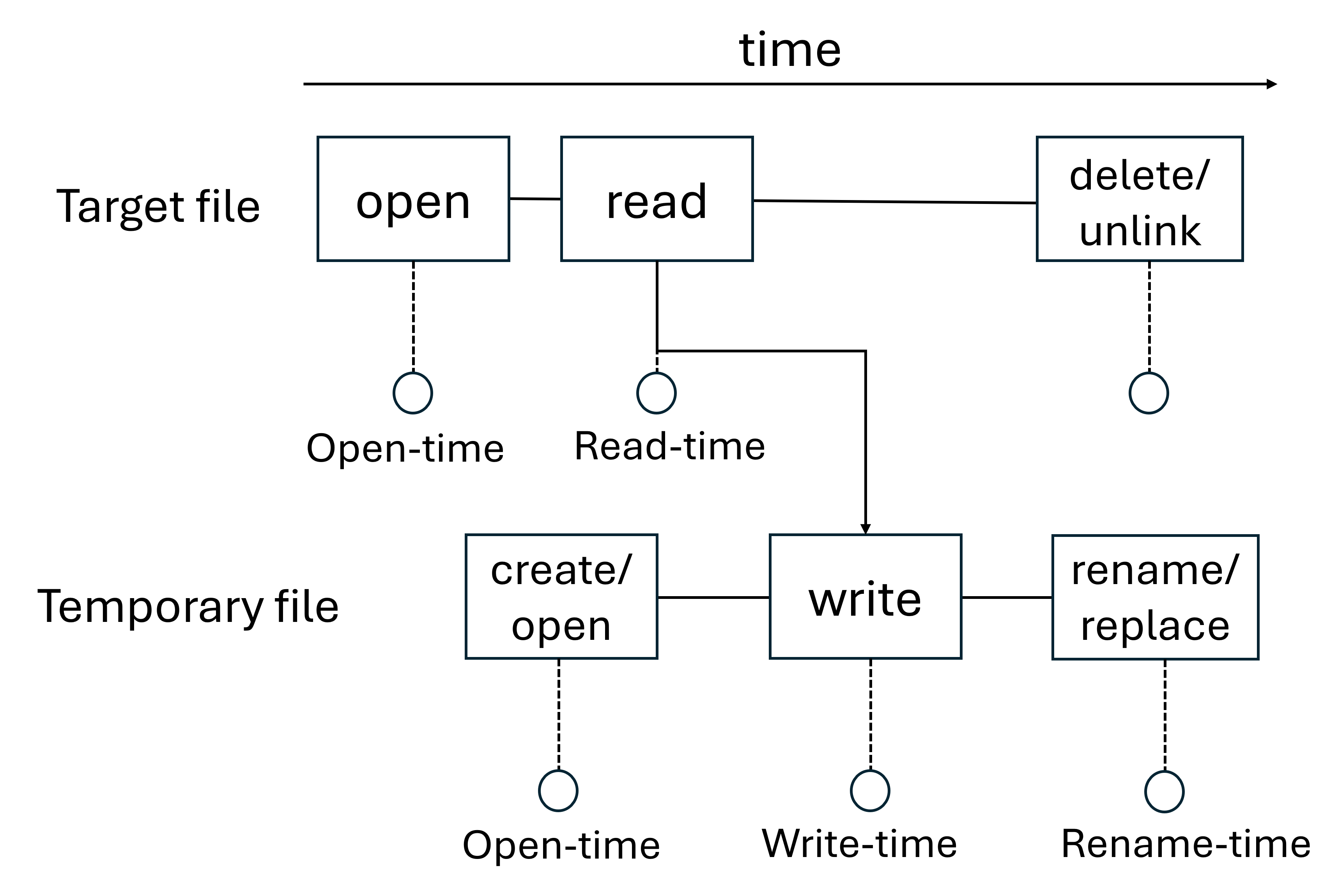}
  \caption{Conceptual model of out-of-place encryption and candidate backup trigger positions.}
  \label{fig:outofplace_encryption}
\end{figure}

Fig.~\ref{fig:outofplace_encryption} shows an out-of-place encryption model.
In this model, ransomware opens and reads an original file, writes the encrypted content to a newly created file, and then deletes or replaces the original file.

In the out-of-place encryption model, the open and read operations may target the original file, whereas the write and rename operations may target a newly created file.
Therefore, even when each operation successfully triggers backup creation, the generated backup may correspond to a different file from the file that is ultimately encrypted.
In particular, a write-triggered backup may target the newly created encrypted file rather than the original file.
Similarly, if a rename operation replaces the original file with the encrypted file, a rename-triggered backup may be too late to preserve the original file's pre-modification content.

These representative models suggest that the selected backup trigger affects when backup creation begins, which file is backed up, and how the backup corresponds to the encrypted file.
We therefore compare four file operations---open, read, write, and rename---as triggers for backup creation.

Our mechanism consists of an eBPF program that monitors file operations in kernel space and a user-space backup process that creates backups of the target files.
Because the experimental environment runs AlmaLinux with XFS, we detect file operations using the XFS-specific kernel functions listed in Table~\ref{tab:xfs_hook_functions}.
We use xfs\_file\_open to detect open operations, xfs\_file\_read\_iter to detect read operations, xfs\_file\_write\_iter to detect write operations, and xfs\_vn\_rename to detect rename operations.
For all four triggers, the backup-side eBPF program attaches a kprobe to the entry point of the corresponding kernel function and obtains information about the target file.

\begin{table}[t]
  \centering
  \caption{XFS kernel functions used for the backup triggers}
  \label{tab:xfs_hook_functions}
  \begin{tabular*}{\columnwidth}{@{\extracolsep{\fill}}lll@{}}
    \hline
    Trigger & Kernel Function & Probe Type \\
    \hline
    Open-time   & xfs\_file\_open        & kprobe \\
    Read-time   & xfs\_file\_read\_iter  & kprobe \\
    Write-time  & xfs\_file\_write\_iter & kprobe \\
    Rename-time & xfs\_vn\_rename        & kprobe \\
    \hline
  \end{tabular*}
\end{table}

When a hooked kernel function is invoked, the eBPF program obtains information identifying the target file and sends the information to the user-space backup process.
The backup process reconstructs the target file path from the received information and determines whether the path is located under the predefined protected directory.
For open, read, and write operations, the path of the file on which the operation is performed is used.
For a rename operation, the source path is used to preserve the file before its name is changed.
If the file is located under the protected directory and has not yet been backed up, the backup process creates a backup of the file.
If multiple events are generated for the same file, no additional backup is created after the first backup has been completed.

Open-time backup issues a backup request when xfs\_file\_open is entered.
Because an open operation generally occurs before read and write operations, this strategy may preserve pre-encryption content at an early stage.
However, it may also create backups of files that are opened only for reading or are not eventually encrypted.

Read-time backup issues a backup request when xfs\_file\_read\_iter is entered.
When ransomware reads the original file before writing encrypted content, the read operation occurs before the destructive write.
Read-time backup may therefore reduce the number of files backed up compared with Open-time backup while still preserving pre-encryption content.
However, ordinary applications may also generate read operations for files that are not subsequently encrypted.
In addition, this strategy cannot create a backup when ransomware modifies a file without first reading it through the monitored function.

Write-time backup issues a backup request when xfs\_file\_write\_iter is entered.
Because a write operation is associated with an actual modification of file content, this strategy may avoid some unnecessary backups generated by open- or read-triggered strategies.
However, the ransomware write continues after the kprobe issues the backup request.
Because the backup itself is performed by the user-space process, the encryption write and backup processing may proceed concurrently.
As a result, part or all of the target file may be modified before the backup process completes, particularly when ransomware encrypts files rapidly.

Rename-time backup issues a backup request when xfs\_vn\_rename is entered.
The source path of the rename operation is used as the backup target.
However, if the rename operation occurs after encryption has been completed, the file content may already be encrypted when the backup request is issued.
In addition, ransomware that overwrites the original file directly or creates a new encrypted file and deletes the original file may not perform a rename operation on the protected file.
In such cases, Rename-time backup does not issue a backup request.

To evaluate each backup trigger independently, only one of the four hooks---open, read, write, or rename---is enabled in each experiment.
The backup program attaches a kprobe only to the kernel function selected by the BACKUP\_TRIGGER configuration.
For example, when BACKUP\_TRIGGER is set to open, only the kprobe attached to xfs\_file\_open is used to trigger backup creation.
All other experimental conditions, including the ransomware detection model, remain unchanged across the experiments.

\section{Experiments}
\label{sec:experiments}

\subsection{Settings and Workloads}
\label{ssec:exp_workload}
Table~\ref{tab:evaluation_env} summarizes the experimental environment.
\begin{table}[H]
  \centering
  \caption{Experimental environment (host and virtual machine)}
  \label{tab:evaluation_env}
  \begin{tabular*}{\columnwidth}{@{\extracolsep{\fill}}ccc@{}}
  \hline
                   & Host machine         & Virtual machine           \\
  \hline
  CPU              & Intel Core i7-12650H & 4 virtual CPU cores       \\
  Memory           & 32 GB (DDR4 2400 MT/s) & 16 GB                    \\
  Storage          & 512 GB (M.2 NVMe)    & 100 GB                     \\
  OS               & Windows 11           & AlmaLinux 10.1             \\
  \hline
  \end{tabular*}
\end{table}

We conduct experiments using samples from multiple ransomware families to evaluate how the choice of backup trigger affects the ability of ROFBS to preserve files during an attack.
The samples evaluated are Sodinokibi, REvil, HelloKitty, AvosLocker, and Conti.
Although Sodinokibi is another name for REvil, we retain the labels assigned at collection to distinguish the two samples.
The samples used in this study were obtained from publicly accessible Internet sources between May and October 2023.
Table~\ref{tab:ransomware} lists the MD5 hashes of the samples used in the experiments.

\begin{table}[H]
  \centering
  \caption{Ransomware samples used in the experiments}
  \label{tab:ransomware}
  \begin{tabular*}{\columnwidth}{@{\extracolsep{\fill}}cc@{}}
    \hline
    Ransomware & MD5 \\
    \hline
    REvil      & 96a157e4c0bef22e0cea1299f88d4745 \\
    HelloKitty & f597d74b84929ded948fc42c6f6210e5 \\
    AvosLocker & d3cafcd46dea26c39dec17ca132e5138 \\
    Conti      & cfb6d21ffe7c4279f761f2351c0810ee \\
    Sodinokibi & c83df66c46bcbc05cd987661882ff061 \\
    \hline
  \end{tabular*}
\end{table}

The protected dataset consists of user files stored under /home/user/victim/.
The backup process is invoked only when access to a path under this directory is observed.

\section{Results}
\label{sec:results}

Tables~\ref{tab:open_results}--\ref{tab:rename_results} present the experimental results for the four backup triggers.
Encrypted Files denotes the number of files encrypted by ransomware.
Backup Files denotes the number of files for which the backup process created backups.
Covered Files denotes the number of encrypted files for which corresponding backup files exist.
The Backup Ratio is calculated using Eq.~\eqref{eq:backup_ratio} as the percentage of Encrypted Files that are Covered Files.
Because backups may also be created for files that are not ultimately encrypted, the number of Backup Files may exceed the number of Encrypted Files.

\begin{equation}
  \mathrm{BackupRatio}
  =
  \frac{N_c}
       {N_e}
  \times 100
  \label{eq:backup_ratio}
\end{equation}

Here, $N_c$ denotes the number of Covered Files, and $N_e$ denotes the number of Encrypted Files.

\begin{table}[H]
  \centering
  \caption{Experimental results for Open-time backup}
  \label{tab:open_results}
  \begin{tabular*}{\columnwidth}{@{\extracolsep{\fill}}lrrrr@{}}
    \hline
    Ransomware & Encrypted & Backup & Covered & Ratio (\%) \\
    \hline
    Conti      & 11 & 12 & 11 & 100.00 \\
    Sodinokibi & 23 &  4 &  3 &  13.04 \\
    AvosLocker &  4 &  9 &  3 &  75.00 \\
    HelloKitty & 15 & 13 &  0 &   0.00 \\
    REvil      & 17 & 28 &  5 &  29.41 \\
    \hline
  \end{tabular*}
\end{table}

Open-time backup achieved Backup Ratios of 100.00\% for Conti, 13.04\% for Sodinokibi, 75.00\% for AvosLocker, 0.00\% for HelloKitty, and 29.41\% for REvil.
For Conti, backups existed for all 11 Encrypted Files.
For AvosLocker, three of the four Encrypted Files were Covered Files.
For HelloKitty, backups were created for 13 files, but none were Covered Files.
For REvil, backups were created for 28 files and covered five Encrypted Files.

\begin{table}[H]
  \centering
  \caption{Experimental results for Read-time backup}
  \label{tab:read_results}
  \begin{tabular*}{\columnwidth}{@{\extracolsep{\fill}}lrrrr@{}}
    \hline
    Ransomware & Encrypted & Backup & Covered & Ratio (\%) \\
    \hline
    Conti      & 18 & 19 & 18 & 100.00 \\
    Sodinokibi & 26 &  4 &  3 &  11.54 \\
    AvosLocker & 54 & 55 &  6 &  11.11 \\
    HelloKitty & 16 &  5 &  0 &   0.00 \\
    REvil      & 24 & 13 &  1 &   4.17 \\
    \hline
  \end{tabular*}
\end{table}

Read-time backup achieved Backup Ratios of 100.00\% for Conti, 11.54\% for Sodinokibi, 11.11\% for AvosLocker, 0.00\% for HelloKitty, and 4.17\% for REvil.
For Conti, backups existed for all 18 Encrypted Files.
For AvosLocker, backups were created for 55 files and covered six Encrypted Files.
For HelloKitty, backups were created for five files, but none were Covered Files.
For REvil, backups were created for 13 files and covered one Encrypted File.

\begin{table}[H]
  \centering
  \caption{Experimental results for Write-time backup}
  \label{tab:write_results}
  \begin{tabular*}{\columnwidth}{@{\extracolsep{\fill}}lrrrr@{}}
    \hline
    Ransomware & Encrypted & Backup & Covered & Ratio (\%) \\
    \hline
    Conti      & 20 & 18 & 17 & 85.00 \\
    Sodinokibi & 27 &  1 &  1 &  3.70 \\
    AvosLocker &  8 &  1 &  1 & 12.50 \\
    HelloKitty & 14 &  9 &  0 &  0.00 \\
    REvil      & 23 &  2 &  1 &  4.35 \\
    \hline
  \end{tabular*}
\end{table}

Write-time backup achieved Backup Ratios of 85.00\% for Conti, 3.70\% for Sodinokibi, 12.50\% for AvosLocker, 0.00\% for HelloKitty, and 4.35\% for REvil.
For Conti, 17 of the 20 Encrypted Files were Covered Files.
For both Sodinokibi and AvosLocker, a backup was created for one file, which was a Covered File.
For HelloKitty, backups were created for nine files, but none were Covered Files.
For REvil, backups were created for two files and covered one Encrypted File.

\begin{table}[H]
  \centering
  \caption{Experimental results for Rename-time backup}
  \label{tab:rename_results}
  \begin{tabular*}{\columnwidth}{@{\extracolsep{\fill}}lrrrr@{}}
    \hline
    Ransomware & Encrypted & Backup & Covered & Ratio (\%) \\
    \hline
    Conti      & 20 &  0 & 0 & 0.00 \\
    Sodinokibi & 39 &  0 & 0 & 0.00 \\
    AvosLocker & 42 &  0 & 0 & 0.00 \\
    HelloKitty & 29 & 13 & 0 & 0.00 \\
    REvil      & 46 &  0 & 0 & 0.00 \\
    \hline
  \end{tabular*}
\end{table}

Rename-time backup yielded a Backup Ratio of 0.00\% for every ransomware sample evaluated.
For Conti, Sodinokibi, AvosLocker, and REvil, no file was backed up.
For HelloKitty, backups were created for 13 files, but none were Covered Files.
\section{Discussion}
\label{sec:discussion}

\subsection{Impact of Backup Triggers on Encrypted Files}
\label{subsec:encrypted_files_discussion}

The same ransomware detection model was used for all backup triggers.
Nevertheless, even for the same ransomware sample, the number of files encrypted by the time of detection varied among the backup triggers.

For Conti, Open-time backup resulted in 11 Encrypted Files, fewer than the 18 observed with Read-time backup and the 20 observed with both Write-time and Rename-time backups.
Similarly, for Sodinokibi, Open-time backup resulted in 23 Encrypted Files, compared with 26 for Read-time backup, 27 for Write-time backup, and 39 for Rename-time backup.
For REvil, Open-time backup also produced the lowest count, with 17 Encrypted Files.
For AvosLocker, Open-time and Write-time backups resulted in four and eight Encrypted Files, respectively, whereas Read-time and Rename-time backups resulted in 54 and 42, respectively.
For HelloKitty, Write-time backup produced the lowest count at 14 Encrypted Files, followed by Open-time backup at 15, Read-time backup at 16, and Rename-time backup at 29.

Because each backup trigger initiates backup processing at a different point, the timing of file reads and storage I/O generated by the backup process also differs.
Open-time backup begins processing before the corresponding read and write operations.
Consequently, backup I/O may have competed with ransomware encryption I/O from an earlier stage and thereby affected the progress of encryption before detection.

In contrast, Rename-time backup created no backups for Conti, Sodinokibi, AvosLocker, or REvil.
Thus, little or no backup-related I/O occurred in these experiments, potentially allowing encryption to progress further than with the other strategies.

\subsection{Impact of Backup Triggers on Backup Ratio}
\label{subsec:backup_ratio_discussion}

For Conti, Open-time and Read-time backups both achieved a Backup Ratio of 100.00\%, whereas Write-time backup achieved 85.00\%.
When open or read operations were used as triggers, backups existed for every encrypted file.
This result suggests that, for the Conti sample used in our experiments, sufficient time was available at the open and read stages to back up files before modification.
The lower Backup Ratio of 85.00\% for Write-time backup suggests that, for some files, the backup process could not complete in time after the backup request was issued upon detection of a write operation.

For Sodinokibi, Open-time, Read-time, and Write-time backups achieved low Backup Ratios of 13.04\%, 11.54\%, and 3.70\%, respectively.
Although Open-time backup outperformed the other strategies, it covered only three files.
Thus, even when open operations were used as triggers, backups corresponding to most encrypted files could not be created for the Sodinokibi sample used in our experiments.
Possible explanations include short intervals between file operations, changes in the correspondence between backup targets and encrypted files, or an encryption procedure that creates separate output files.

For AvosLocker, Open-time backup achieved a Backup Ratio of 75.00\%, exceeding the 11.11\% achieved by Read-time backup and the 12.50\% achieved by Write-time backup.
With Open-time backup, three of the four Encrypted Files were Covered Files.
In contrast, Read-time backup created backups for 55 files, but only six of the 54 Encrypted Files were Covered Files.
This result indicates that, for the AvosLocker sample used in our experiments, triggering on read events created many backups, but most did not protect files that were ultimately encrypted.

For HelloKitty, every backup trigger yielded a Backup Ratio of 0.00\%.
Although all four strategies---Open-time backup, Read-time backup, Write-time backup, and Rename-time backup---created Backup Files, none were Covered Files.
This result shows that detecting events and creating backups alone may not ensure that the resulting backups correspond to the files ultimately encrypted.
For the HelloKitty sample used in our experiments, the relationship between paths or files may have changed during encryption, or the ransomware may have generated and encrypted files other than the backup targets.

For REvil, Open-time, Read-time, and Write-time backups achieved Backup Ratios of 29.41\%, 4.17\%, and 4.35\%, respectively.
Open-time backup created backups for 28 files, but only five of the 17 Encrypted Files were Covered Files.
Read-time and Write-time backups each covered only one file.
Thus, for the REvil sample used in our experiments, triggering on open operations was more effective than triggering on read or write operations, but open operations alone did not protect most encrypted files.

Rename-time backup yielded a Backup Ratio of 0.00\% for every ransomware sample.
Because no Backup Files were created for Conti, Sodinokibi, AvosLocker, or REvil, these samples may not have renamed protected files, or the system may not have observed rename events that it could identify as backup targets.
For HelloKitty, backups were created for 13 files, but none were Covered Files.
Thus, although HelloKitty generated rename events, the resulting backups did not correspond to the files ultimately classified as encrypted.
Accordingly, rename operations did not function adequately as a standalone backup trigger for the samples evaluated in this study.

Overall, the effectiveness of a backup trigger depends on the sequence of file operations performed by the ransomware.
Open and read operations were effective triggers for Conti, whereas open operations yielded higher Backup Ratios than the other triggers for AvosLocker, Sodinokibi, and REvil.
In contrast, none of the triggers produced any Covered Files for HelloKitty.
These findings indicate that no single file operation served as an optimal backup trigger across the ransomware samples evaluated in this study.

\subsection{Relationship between Backup Files and Covered Files}
\label{subsec:backup_covered_discussion}

Backup Files denotes the number of files for which backups were created; however, increasing the number of Backup Files does not necessarily increase the number of Covered Files.

For Conti, Open-time backup produced 12 Backup Files and 11 Covered Files, whereas Read-time backup produced 19 Backup Files and 18 Covered Files.
Thus, most backups triggered by open and read operations for Conti corresponded to files that were subsequently encrypted.

In contrast, Read-time backup for AvosLocker produced 55 Backup Files but only six Covered Files.
Open-time backup for REvil produced 28 Backup Files but only five Covered Files.
For HelloKitty, every strategy created Backup Files, yet none were Covered Files.
These results demonstrate that even a large number of backups does not improve the Backup Ratio unless those backups correspond to the files ultimately encrypted.

For Write-time backup, both Sodinokibi and AvosLocker produced one Backup File and one Covered File.
However, Covered Files accounted for only a small proportion of Encrypted Files.
This finding suggests that backups triggered by write operations may correspond to encrypted files, but the strategy may create backups for only a limited subset of the targets.

Therefore, evaluating a backup trigger requires examining not only its Backup Ratio but also the relationship between Backup Files and Covered Files.

\subsection{Limitations}
\label{subsec:limitations}

Our evaluation is limited to XFS on AlmaLinux.
The results therefore may not generalize to other Linux distributions, kernel versions, or file systems.

Moreover, the effectiveness of a backup trigger may vary among sample versions and configurations, even within the same ransomware family.
The results should therefore be interpreted as observations specific to the samples used in this study.

Finally, although the same ransomware detection model was used for every strategy, the timing and volume of I/O generated by the backup process varied among the triggers.
Backup processing may therefore have affected both the progress of encryption and the timing of detection.
Future work should record backup I/O and ransomware I/O separately to isolate and evaluate the effect of the backup trigger itself.
\section{Conclusion}
\label{sec:conclusion}

In this study, we compared four file operations---open, read, write, and rename---as triggers for backup creation in ROFBS.
We implemented each strategy by attaching an eBPF program to the corresponding XFS kernel function.
In experiments with Conti, Sodinokibi, AvosLocker, REvil, and HelloKitty, the mean Backup Ratios across the five samples were 43.49\% for Open-time backup, 25.36\% for Read-time backup, 21.11\% for Write-time backup, and 0.00\% for Rename-time backup.
Among the four strategies compared, Open-time backup thus achieved the highest mean Backup Ratio.

However, the effectiveness of each strategy varied among ransomware samples.
For Conti, Open-time and Read-time backups achieved Backup Ratios of 100.00\%, and Write-time backup achieved 85.00\%; for HelloKitty, all strategies yielded 0.00\%.
Read-time backup created many backups, but this increase did not translate into a corresponding increase in Covered Files.
Write-time backup created relatively few backups and yielded low Backup Ratios for every sample except Conti.
Rename-time backup produced no Covered Files for any sample evaluated.

These findings demonstrate that the file operation selected as the backup trigger is an important design factor that affects the Backup Ratio, the number of backups created, and the number of files encrypted before detection.
Within the scope of our experiments, open operations were the most effective trigger; however, no single trigger provided sufficient protection against all ransomware samples.
Future work will analyze the file-operation sequences of individual ransomware families in detail and investigate backup-trigger strategies that combine multiple file operations.

\section*{Acknowledgment}
GPT-5.6 Sol was used for Japanese-to-English translation, grammar checking, and editorial purposes throughout this manuscript.
All outputs were inspected and revised by the authors to ensure accuracy and originality.

\bibliographystyle{unsrt}
\bibliography{open}

\end{document}